\documentclass[nofootinbib,notitlepage,aps,prl,reprint,superscriptaddress]{revtex4-1}

\renewcommand{\thesection}{\Roman{section}}

\renewcommand{\figurename}{FIG.}

\usepackage{mathtools}
\usepackage[english]{babel}
\usepackage[utf8]{inputenc}
\usepackage{amsmath}
\usepackage{amsfonts}
\usepackage{amssymb}
\usepackage{graphicx}
\usepackage{bbold}
\usepackage{epstopdf}
\usepackage{hyperref}
 
\renewcommand*{\Im}{\operatorname{Im}} 

\DeclarePairedDelimiter\ket{\lvert}{\rangle}

\newcommand{\appropto}{\mathrel{\vcenter{
  \offinterlineskip\halign{\hfil$##$\cr
    \propto\cr\noalign{\kern2pt}\sim\cr\noalign{\kern-2pt}}}}}

\begin{document}
\renewcommand{\figurename}{FIG.}
\renewcommand{\tablename}{TABLE}
\title{Unconditional preparation of squeezed vacuum from Rabi interactions}
\author{Jacob Hastrup}
\email{jhast@fysik.dtu.dk}
\affiliation{Center for Macroscopic Quantum States (bigQ), Department of Physics, Technical University of Denmark, Building 307, Fysikvej, 2800 Kgs. Lyngby, Denmark}
\author{Kimin Park}
\email{kimpa@fysik.dtu.dk}
\affiliation{Center for Macroscopic Quantum States (bigQ), Department of Physics, Technical University of Denmark, Building 307, Fysikvej, 2800 Kgs. Lyngby, Denmark}
\affiliation{Department of Optics, Palacky Univeristy, 77146 Olomouc, Czech Republic}
\author{Radim Filip}
\email{filip@optics.upol.cz}
\affiliation{Department of Optics, Palacky Univeristy, 77146 Olomouc, Czech Republic}
\author{Ulrik Lund Andersen}
\email{ulrik.andersen@fysik.dtu.dk}
\affiliation{Center for Macroscopic Quantum States (bigQ), Department of Physics, Technical University of Denmark, Building 307, Fysikvej, 2800 Kgs. Lyngby, Denmark}

\begin{abstract}
Squeezed states of harmonic oscillators are a central resource for continuous-variable quantum sensing, computation and communication. Here we propose a method for the generation of very good approximations to highly squeezed vacuum states with low excess anti-squeezing using only a few oscillator-qubit coupling gates through a Rabi-type interaction Hamiltonian. This interaction can be implemented with several different methods, which has previously been demonstrated in superconducting circuit and trapped-ion platforms. The protocol is compatible with other protocols manipulating quantum harmonic oscillators, thus facilitating scalable continuous-variable fault-tolerant quantum computation. 
\end{abstract}
\date{\today}

\maketitle

\section{INTRODUCTION}
Quantum continuous variables have become an increasingly promising platform for quantum information processing \cite{braunstein2005quantum}. In particular, extraordinary experimental progress has been made over the last few years in trapped-ion and superconducting circuit platforms towards fault-tolerant quantum computation \cite{ofek2016extending,hu2019quantum,campagne2020quantum}. One of the most promising routes towards fault-tolerant continuous-variable quantum computation is the Gottesman-Kitaev-Preskill encoding \cite{gottesman2001encoding}, which has gained substantial interest over the past few years due to experimental and theoretical developments \cite{fluhmann2019encoding,campagne2020quantum,terhal2020towards}. For this encoding, highly squeezed states are an important resource for constructing high-quality states \cite{terhal2016encoding,fluhmann2019encoding,hastrup2019measurement}. The current record for squeezed vacuum is 15 dB \cite{vahlbruch2016detection} in an optical field using a parametric amplifier. However, non-Gaussian operations are difficult to realise efficiently in the optical regime, and thus it is challenging with current technology to further utilize this highly squeezed state for quantum computation. 

On the other hand, qubit-coupled oscillators, such as a motional state of a trapped ion or a microwave cavity field coupled to a superconducting qubit can be manipulated with non-Gaussian operations via the qubit ancilla. In fact, universal control of the harmonic oscillator is in principle possible in such systems \cite{law1996arbitrary,krastanov2015universal}, although many protocols, such as squeezed state preparation, require specialized methods to be efficient. 12.6 dB squeezing has been reported in the motional state of a trapped ion \cite{kienzler2015quantum} using a reservoir engineering technique \cite{cirac1993dark}. This technique has the advantage of achieving squeezing in a steady-state configuration, thus facilitating the experimental implementation. However, the process utilizes spontaneous relaxation processes, the rates of which limits the speed at which the state is created and thus ultimately the achievable squeezing due to dephasing during the protocol. In the microwave regime 10 dB squeezing has been experimentally demonstrated \cite{castellanos2008amplification}. This was achieved using a parametric amplifier realised by a metamaterial consisting of multiple Josephson junctions. 

%Note that in the trapped-ion experiment, the squeezing was not directly measured, but estimated based on the phonon number distribution, and thus actual squeezing was like a few dB lower.

Here we propose a method for the preparation of an approximate squeezed vacuum state in an oscillator strongly coupled to a qubit using only a few unitary interactions through the Rabi Hamiltonian \cite{forn2019ultrastrong,kockum2019ultrastrong}. This Hamiltonian has been experimentally demonstrated in trapped-ions and superconducting circuits \cite{fluhmann2018sequential,lv2018quantum,langford2017experimentally,campagne2020quantum}, and plays a key role in the Gottesman-Kitaev-Preskill encoding scheme of these platforms \cite{campagne2020quantum,fluhmann2019encoding,terhal2020towards}. Thus the protocol facilitates the generation of highly squeezed states using a method that is compatible with further manipulation of the oscillator. Our protocol for the generation of squeezed vacuum is radically different from the common approach based on parametric amplification, and represents a fundamentally new view on squeezed vacuum generation. The obtainable amount of squeezing depends on the types and magnitude of noise in the particular system, but can generally be improved through faster interactions, e.g. through an increased power of the driving fields which control the interaction. Furthermore, the achievable amount of Fisher information is particularly robust against qubit errors during the protocol, making the generated states useful for sensing applications \cite{giovannetti2004quantum,ivanov2018quantum,penasa2016measurement,ivanov2016high,hempel2013entanglement,burd2019quantum}. In particular, squeezed states can be used to detect displacements in the considered platforms using either the qubit-coupling \cite{duivenvoorden2017single} or homodyne detection \cite{naghiloo2020heat}. Finally, squeezed states serve as a fundamental resource for continuous variable communication \cite{braunstein2005quantum} which could find applications facilitating short-range connections in microwave circuits \cite{pfaff2017controlled}.

\section{PROTOCOL}

We consider a quantum harmonic oscillator described by the quadrature operators $\hat{X}$ and $\hat{P}$ satisfying $[\hat{X},\hat{P}]=i$. In the vacuum state the oscillator exhibits equal fluctuations in each quadrature of magnitude $\langle\hat{X}^2\rangle=\langle\hat{P}^2\rangle=\frac{1}{2}$. The aim of the protocol is to generate a state squeezed in the $P$-quadrature by a relative amount $\Delta^2<1$ such that $\langle\hat{P}^2\rangle=\frac{\Delta^2}{2}$. The protocol can straightforwardly be generalized to generate squeezing along an arbitrary quadrature direction, either by a rotation from the natural evolution of an the squeezed state under a harmonic potential, or by a suitable change of quadrature operators during the protocol. It is common to quantify the squeezing in dB relative to the vacuum as $\Delta_\textrm{dB}=-10\log_{10}(\Delta^2)$. A pure squeezed vacuum state, $\ket{\textrm{sqvac}}_\Delta$, can be written in the $P$-, $X$- and coherent state bases as:
%\begin{linenomath*}
\begin{subequations}
\begin{align}
    \ket{\textrm{sqvac}}_{\Delta}&\propto\int dp \exp\left(-\frac{p^2}{2\Delta^2}\right)\ket{p}\\ 
    &\propto\int dx \exp\left(-\frac{x^2}{2\Delta^{-2}}\right)\ket{x}\\ 
    &\propto\int d\alpha\exp\left(-\frac{\alpha^2}{\Delta^{-2}-1}\right)\ket{\alpha}, \label{eq:coherent}
\end{align}
\end{subequations}
%\end{linenomath*}
where the last line is only valid for $\Delta<1$ and the integral is over real $\alpha$. The coherent states with real $\alpha$ are defined as $\ket{\alpha} = e^{-i\sqrt{2}\alpha\hat{P}}\ket{\textrm{vac}}\propto \int dx \exp\left(-\left(x-\sqrt{2}\alpha\right)^2/2\right)\ket{x}$.
For our approach, it is useful to view the squeezed state in the coherent state basis, as our strategy will be to directly construct a superposition of coherent states which resembles \eqref{eq:coherent}. Since the coherent states form an overcomplete basis with large overlap between close-lying states, we can expect that equation \eqref{eq:coherent} holds to a good approximation even if we discretize the integral:
%\begin{linenomath*}
\begin{equation}
\ket{\textrm{sqvac}}_\Delta \appropto \sum_{\alpha_s\in \mathfrak{L}} \exp\left(-\frac{\alpha_s^2}{\Delta^{-2}-1}\right)\ket{\alpha_s},
\label{eq:approx}
\end{equation}
%\end{linenomath*}
where $\mathfrak{L}$ is a lattice on the real line. While the right hand side of Eq.\ \eqref{eq:approx} is technically a non-Gaussian state, we find that Eq.\ \eqref{eq:approx} is a good approximation for a lattice spacing of up to $\sim 1.5$, which is on the order of the width of a coherent state (details are presented in the Supplementary Material, \ref{app:squeezedApprox}). It is therefore possible to construct a highly squeezed state, practically indistinguishable from a Gaussian squeezed state, from a relatively sparse superposition of coherent states. A probabilistic method based on this approach was proposed in \cite{roszak2015decoherence}.

\begin{figure}
\centering
\includegraphics{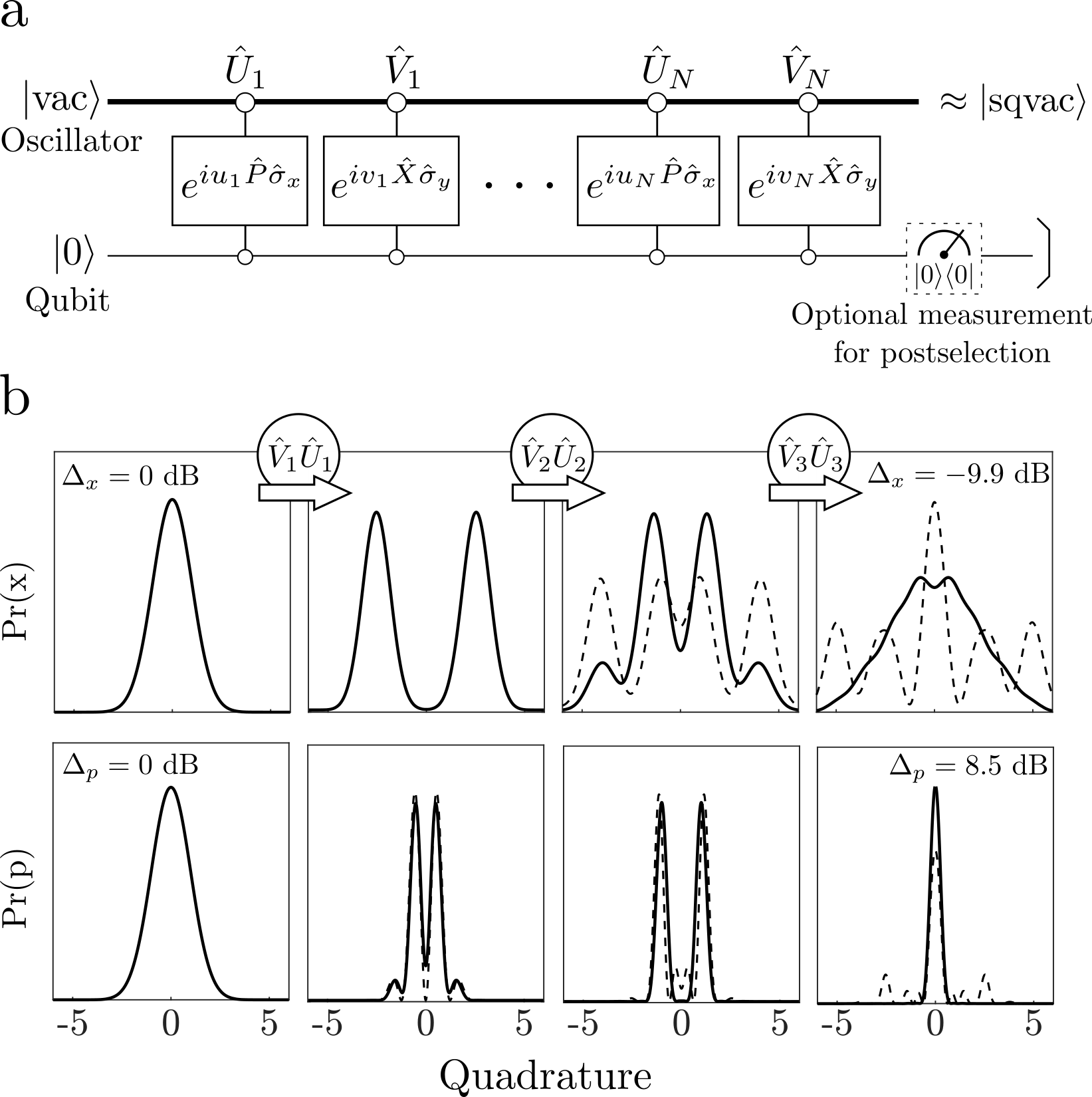}
\caption{(a) Circuit diagram for the generation of a $P$-squeezed vacuum state. The protocol consists of $N$ steps of interactions through the Hamiltonians $\hat{P}\hat{\sigma}_x$ and $\hat{X}\hat{\sigma}_y$. The interaction parameters $u_k$ and $v_k$ varies from step to step and are found numerically to optimize the protocol. The protocol is deterministic, but the performance can be slightly improved by measuring the qubit and postselecting on the outcome 0. (b) Evolution of the quadrature distributions during each step of the protocol. Dashed lines correspond to interaction parameters given by Eq. \eqref{eq:parameters} with $L=0.45$ and solid lines correspond to numerically optimized interaction parameters.}
\label{fig:scheme}
\end{figure}

We now present a deterministic method to efficiently construct such a superposition of coherent states using a qubit ancilla. The circuit diagram of the protocol is shown in Fig. \ref{fig:scheme}a. We use two Rabi-type interaction Hamiltonians $\hat{P}\hat{\sigma}_x$ and $\hat{X}\hat{\sigma}_y$ \cite{forn2019ultrastrong,kockum2019ultrastrong}, where $\hat{\sigma}_x$ and $\hat{\sigma}_y$ are the Pauli $x$ and $y$ operators of the qubit. Such Hamiltonians can be efficiently implemented \cite{haljan2005spin,ballester2012quantum,mezzacapo2014digital,campagne2020quantum}, e.g. using a two-tone drive which has been experimentally demonstrated in trapped-ions \cite{fluhmann2018sequential,lv2018quantum} or from a dispersive Jaynes-Cumming Hamiltonian which has been demonstrated in superconducting circuits \cite{campagne2020quantum,langford2017experimentally}. The protocol consists of $N$ pairs of different interactions. The first type of interaction, $\hat{U}_k=\exp(iu_k\hat{P}\hat{\sigma}_x)$, displaces the oscillator in a direction depending on the state of the qubit, while the following interaction, $\hat{V}_k=\exp(iv_k\hat{X}\hat{\sigma}_y)$, approximately disentangles the qubit and the oscillator. The repeated application of these interactions creates a superposition of $2^N$ coherent states, and leaves the qubit back in the ground state. Note that the interactions used are conceptually similar to the protocol in Ref. \cite{hastrup2019measurement} which aims to generate a grid state starting from squeezed vacuum. However, unlike this previous work, our target state consists of overlapping coherent states, for which the approximations used in Ref.\ \cite{hastrup2019measurement} do not hold. In this work, we show that this limitation can be overcome surprisingly well by numerically optimizing the interaction parameters, $u_k$ and $v_k$, for each step, which enables the generation of squeezed states using Rabi interactions. Still, to understand why the protocol works and to have a good initial guess from which to optimize the interaction parameters, it is instructive to consider a specific set of interaction parameters given by
%\begin{linenomath*}
\begin{subequations}
\begin{align}
u_k &=\begin{cases}
                2^{N-1}\sqrt{2}L, & \text{if $k=1$,}\\
               -2^{N-k}\sqrt{2}L & \text{if $k>1$,}\\
            \end{cases} \\
            v_k &=\begin{cases}
               2^{-(N-k)}\frac{\pi}{4\sqrt{2}L}, & \text{if $k<N$,}\\
               -\frac{\pi}{4\sqrt{2}L} & \text{if $k=N$,}\\
            \end{cases}
\end{align}
\label{eq:parameters}
\end{subequations}
%\end{linenomath*}
where $L$ is a free parameter, which determines the spacing of the resulting grid of coherent states. Each step aims to double the number of coherent states in the superposition. The first interaction, $\hat{U}_1$, displaces the oscillator and entangles it with the qubit:
%\begin{linenomath*}
\begin{align}
    \hat{U}_1\ket{\textrm{vac}}\ket{0} = \ket{-2^{N-1}L}\ket{+} + \ket{2^{N-1}L}\ket{-}.
\end{align}
%\end{linenomath*}
Measuring the qubit in the in the $\ket{0} = \left(\ket{+}+\ket{-}\right)/\sqrt{2}$ or $\ket{1}= \left(\ket{+}-\ket{-}\right)/\sqrt{2}$ state leaves the oscillator in a superposition of two coherent states, known as a Schrödingers cat state, which has been experimentally demonstrated using exactly this type of interaction \cite{langford2017experimentally}. In our protocol, however, we do not require qubit measurements to disentangle the qubit and the oscillator. Instead, we apply a second interaction, $\hat{V}_1$, which approximately disentangles the qubit and the oscillator,
\begin{widetext}
\begin{align}
    \hat{V}_1\hat{U}_1\ket{\textrm{vac}}\ket{0} & = \hat{V}_1\left(\ket{-2^{N-1}L}\ket{+} + \ket{2^{N-1}L}\ket{-}\right)\nonumber\\
    &=\frac{e^{-i\pi/4}}{2}\Bigg(\left[e^{-i\pi/8}\ket{-2^{N-1}L+i\sqrt{2}v_1} + ie^{i\pi/8}\ket{2^{N-1}L + i\sqrt{2}v_1}\right]\ket{+i}\nonumber\\
    &+\left[ie^{i\pi/8}\ket{-2^{N-1}L-i\sqrt{2}v_1} + e^{-i\pi/8} \ket{2^{N-1}L - i\sqrt{2}v_1}\right]\ket{-i}\Bigg)\nonumber\\
    &= \left(\ket{2^{N-1}L} - \ket{-2^{N-1}L}\right)\ket{1} +\mathcal{O}(v_1).
\end{align}
\end{widetext}
where $\ket{\pm i}=(\ket{0}\pm i\ket{1})$ are the $\hat{\sigma}_y$ eigenstates and we have used the relation $\langle\beta|\alpha\rangle=e^{i\Im(\beta^*\alpha)}e^{-|\beta-\alpha|^2/2}$ to write $\ket{-2^{N-1}L \pm i\sqrt{2}v_1} = e^{\pm i\pi/8}\ket{-2^{N-1}L} + \mathcal{O}(v_1)$, where $\mathcal{O}(v_1)$ denotes terms on the order $v_1$, which can be neglected when the coherent states are well separated, i.e. when $2^{N}L\gg 1$. Note that due to the complete absence of a measurement, the method does not rely on neither postselection or active feed-forward. Moreover, we circumvent accumulated measurement-induced noise such as measurement errors and bosonic noise. Each subsequent pair of interactions splits each coherent state into two, doubling the total number of peaks. Thus after all $N$ steps we produce the state:
\begin{equation}
\prod_k^N\hat{V}_k\hat{U}_k\ket{\textrm{vac}}\ket{0} \approx \left(\sum_{s=0}^{2^N-1}\ket{(2s+1-2^N)L}\right)\ket{0}, \label{eq:result}
\end{equation}
i.e. the oscillator ends in a superposition of multiple coherent states, similar to the target state of Eq. \eqref{eq:approx}. However, there are two main issues with Eq. \eqref{eq:result}: First, the result yields an equal superposition of the coherent states, whereas our target state is convolved with a Gaussian envelope. Secondly, the approximation of Eq.  \eqref{eq:result} is only valid when the coherent states are sufficiently well separated, but to obtain a good approximation to a squeezed state, the coherent states need to be overlapping. It turns out that one can overcome these issues surprisingly well by tuning the interaction strengths of the protocol. This is illustrated in Fig. \ref{fig:scheme}b, showing the quadrature distributions of the oscillator for each step of the protocol. The solid lines represent the distributions using numerically optimized parameters while the dashed lines show the result for the parameters given by Eq. \eqref{eq:parameters}. Using the parameters of Eq. \eqref{eq:parameters}, the final P-quadrature distribution has side-lopes which effectively reduces $\langle \hat{P}^2\rangle$ to that of vacuum. For the numerically optimized parameters, however, these lopes vanish, thus yielding a highly squeezed state. 
There are multiple reasons why the protocol is improved by tuning the interaction parameters. Firstly, by tuning the strengths of the second interaction, $v_k$, the qubit and the oscillator do not completely disentangle, so the subsequent controlled displacement, $\hat{U}_{k+1}$, does not split each peak equally, but with a preferred direction, resulting in an unequal final distribution. This enables us to obtain an approximately Gaussian envelope over the resulting superposition as in Eq. \eqref{eq:approx}. Additionally, as the states start to overlap, the disentangling interactions, $\hat{V}_k$, have to be adjusted to optimize the disentanglement between the oscillator and qubit in the final step. Furthermore, when the coherent states of unequal amplitude are overlapping their peaks are effectively slightly shifted, which can be corrected by tuning the displacement interactions, $\hat{U}_{k}$. 

In figure \ref{fig:scheme}b we have chosen the interaction parameters to optimize only the squeezing in the P quadrature, yielding $\Delta_p=8.5$ dB for $N=3$. The anti-squeezing in the X quadrature is slightly in excess, $\Delta_x=-9.9$ dB, due to the underlying non-Gaussianity of the state, but still quite close to the transform-limit, showing that the output is a good approximation to a pure squeezed vacuum state. Still, even lower anti-squeezing is possible without significantly compromising the squeezing by choosing the interaction parameters differently, e.g. one can obtain $\Delta_p=8.0$ dB with $\Delta_x=-8.3$ dB (see Supplementary Material).

\begin{figure}[t]
\centering
\includegraphics{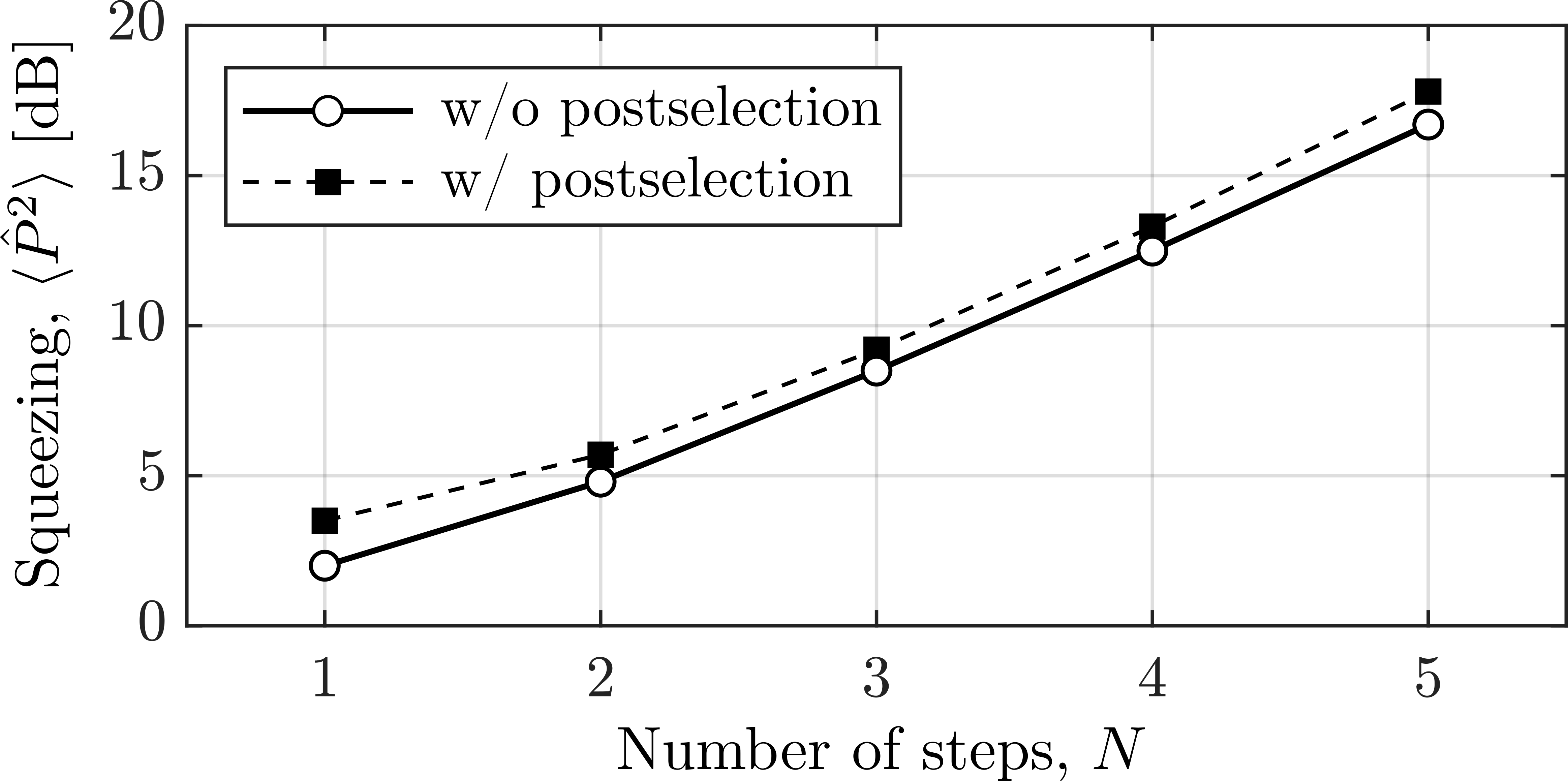}
\caption{Resulting squeezing as a function of the number of steps of the protocol. For each step, the squeezing increases with $\sim3$-$4$ dB. The dashed curve corresponds to adding the optional measurement in Fig. \ref{fig:scheme}a and post-selecting on the outcome $\ket{0}$, which occurs with high probability.}
\label{fig:result}
\end{figure}

The resulting squeezing for the numerically optimized parameters is shown by the circles in Fig. \ref{fig:result}. Only a few number of steps is required to generate a highly squeezed state, which is expected as the number of coherent states in the superposition increases exponentially with $N$.
It is possible to further improve this result by roughly 1 dB by post-selecting states for which the qubit is measured in the $\ket{0}$ state after all interactions. The protocol should leave the qubit in state $\ket{0}$ according to Eq. \eqref{eq:result}, but since this is an approximate result, a projection onto $\ket{0}$ helps improving this approximation. A post-selectable result therefore also occurs with high probability. Note that while the produced states are fundamentally non-Gaussian, due to the finite and discrete number of underlying coherent states, the output is practically indistinguishable from a Gaussian squeezed stated. We confirm this in the Supplementary Material, showing that the generated states have fidelities of $>0.99$ with respect to pure Gaussian squeezed vacuum states.

\begin{figure}[t]
\centering
\includegraphics{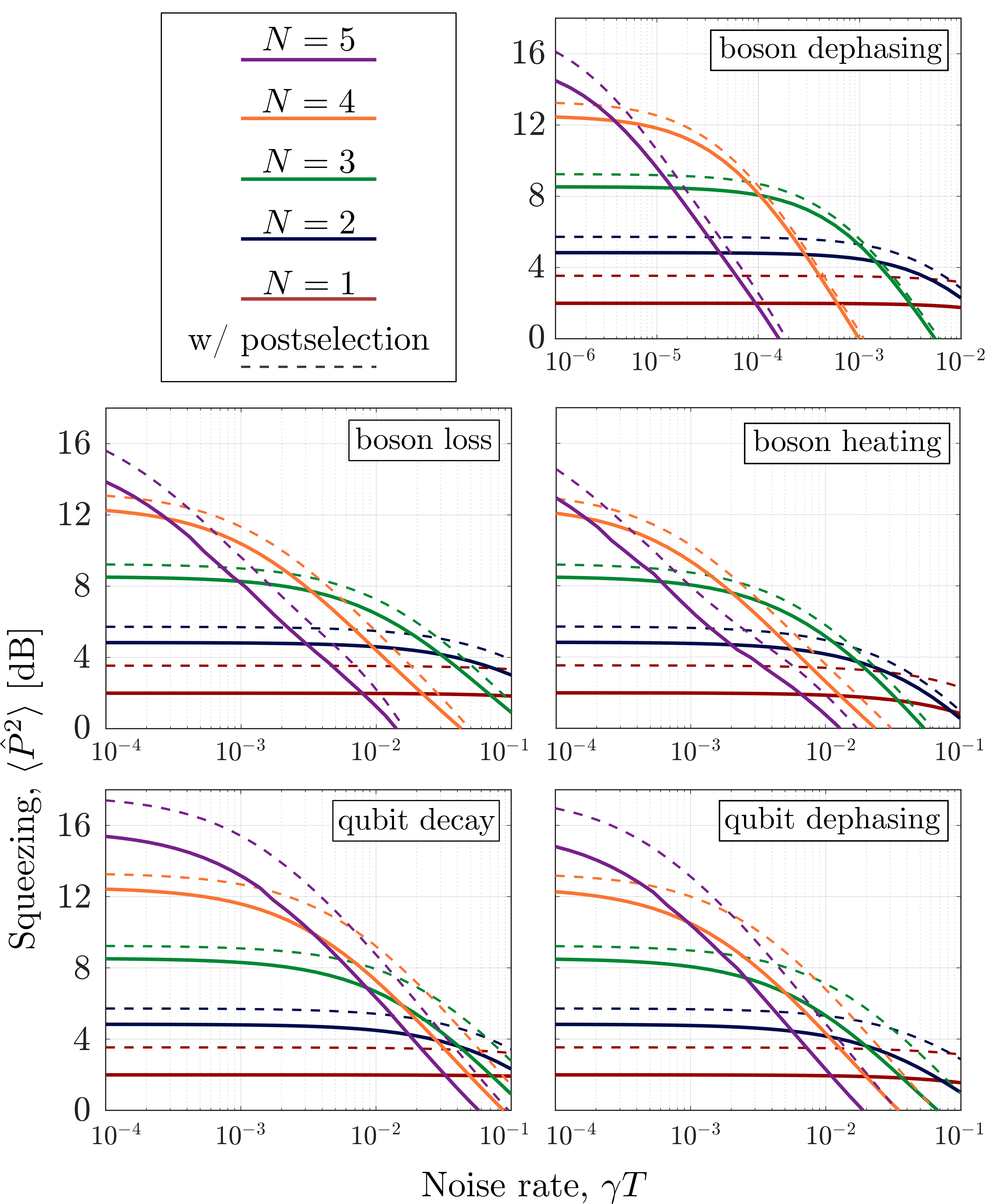}
\caption{Obtainable squeezing as a function of noise rate for different noise sources. For each noise source and noise rate, there exists an optimal number of steps. Post-selecting on the outcome $\ket{0}$ can in some case improve the performance by more than 2 dB.}
\label{fig:noise}
\end{figure}

From Fig. \ref{fig:result} we see that increasing the number of interactions monotonically increases the resulting squeezing. Thus the protocol can fundamentally be scaled to achieve large amounts of squeezing. However, real physical systems are affected by noise, such as dephasing and loss, which will accumulate during the protocol. Assuming the time for each interaction is proportional to the absolute interaction parameter, the total protocol duration roughly doubles each time $N$ is augmented, as the interaction parameters approximately scale as $2^N$ according to Eqs. \eqref{eq:parameters}. The increased squeezing therefore eventually gets counteracted by the accumulated noise. To study the effects of noise, we simulate the protocol using the Master equation,
\begin{equation}
    \frac{\textrm{d}\rho}{\textrm{d}t}=-\frac{i}{\hbar}[\hat{H},\rho] + \hat{L}\rho\hat{L}^\dagger - \frac{1}{2}\left(\hat{L}^\dagger\hat{L}\rho+\rho\hat{L}^\dagger\hat{L}\right),
    \label{eq:master}
\end{equation}
where $\rho$ is the density matrix of the composite boson-qubit system. The Hamiltonian, $\hat{H}$, is $\pm\frac{\hbar}{T}\hat{P}\hat{\sigma}_x$ or $\pm\frac{\hbar}{T}\hat{X}\hat{\sigma}_y$ during the two types of interactions, where the sign depends on the sign of the interaction parameter and $T$ is a timescale denoting the time required to implement $\exp\left(i\hat{P}\hat{\sigma}_x\right)$ or $\exp\left(i\hat{X}\hat{\sigma}_y\right)$. Thus the first interaction takes place in a time $Tu_1$ after which the interaction Hamiltonian abruptly changes to the next one. $\hat{L}$ is the Lindblad noise operator, which determines the type of noise. We consider five types:
\begin{itemize}
\item Boson loss: $\hat{L}=\sqrt{\gamma}\hat{a}$
\item Boson dephasing: $\hat{L}=\sqrt{\gamma}(\hat{a}\hat{a}^\dagger + \hat{a}^\dagger\hat{a})$
\item Boson heating\footnote{Boson heating is described by two Lindblad operators with strengths dependent on the temperature of the environment bath, $\hat{L}_1=\sqrt{\gamma_c(1+\bar{n})}\hat{a}$ and $\hat{L}_2 =\sqrt{\gamma_c\bar{n}}\hat{a}^\dagger$, where $\gamma_c$ is the coupling rate to the bath with mean occupation number $\bar{n}$. Here we define the heating rate $\gamma_c\bar{n}\equiv\gamma$ and assume $\bar{n}\gg1$ such that $\gamma_c(1+\bar{n})\approx \gamma_c\bar{n}$ to isolate the effect of heating rather than thermalization.}: $\hat{L}_1=\sqrt{\gamma}\hat{a}, \hat{L}_2 =\sqrt{\gamma}\hat{a}^\dagger$
\item Qubit decay: $\hat{L}=\sqrt{\gamma}(\hat{\sigma}_x + i\hat{\sigma}_y)/2$
\item Qubit dephasing: $\hat{L}=\sqrt{\gamma}\hat{\sigma}_z$\\
\end{itemize} 
where $\gamma$ is the noise rate and $\hat{a}=\left(\hat{X}+i\hat{P}\right)/\sqrt{2}$ is the bosonic annihilation operator. The results are shown in Fig. \ref{fig:noise}. For each noise source and noise rate we find that there exists an optimum number of interactions. Bosonic noise is seen to have a bigger impact compared to qubit noise. Especially boson dephasing can heavily reduce the obtained squeezing, which is expected as squeezed states are generally sensitive to dephasing. The dashed lines show the outcomes which are post-selected on measuring the qubit in state $\ket{0}$. We observe that the positive effect of post-selection is now slightly larger compared to Fig. \ref{fig:result}, especially for qubit-associated noise. This is because the noise can result in the qubit ending up in the $\ket{1}$ state, in which case the presence of noise can be detected and the event discarded. The post-selection strategy can therefore effectively reduce the effect of noise. 

A key property of squeezed states is their ability to detect displacements \cite{giovannetti2004quantum,ivanov2018quantum,penasa2016measurement}, which is quantified by the Fisher information \cite{paris2004quantum}, $I_C$. While the quadrature squeezing is affected by all noise types, the Fisher information turns out to be quite robust against qubit errors, as we show in the Supplementary Material. For example, for $N=4$ with a qubit decay rate of $\gamma T=7\times 10^{-1}$ we calculate $I_C=56$, which is equivalent to that of an $11.5$ dB squeezed vacuum state. Thus the generated states can still be useful for sensing applications, even though they are generated under noisy conditions.

Finally, we benchmark our approach using noise figures from two recent experiments which implement exactly the types of interactions needed for our protocol in trapped ions \cite{fluhmann2019encoding} and microwave cavities \cite{campagne2020quantum}. For the parameters in \cite{fluhmann2019encoding} we calculate 9.3 dB squeezing and a Fisher information of $I_C=63$ (equivalent to a 11.9 dB squeezed state) using $N=4$. While this is slightly lower than the 12.6 dB reported in the trapped ion experiment in Ref. \cite{kienzler2015quantum}, we point out that in their experiment the quadrature squeezing was not measured directly, but estimated using only the phonon population distribution. Since the quadrature squeezing is sensitive to small fluctuations and the coherence of the state, the 12.6 dB is likely overestimating the actual squeezing of the generated state. For the parameters in \cite{campagne2020quantum} we obtain an optimum squeezing of 7.0 dB at $N=3$, and an optimum Fisher information at of $I_C=86$ (equivalent to a 13.3 dB squeezed state) at $N=5$. The high Fisher information relative to the quadrature squeezing is due to the effect of qubit errors. In particular, these errors translate into non-Gaussian features of the output state which primarily degrade the quadrature squeezing (see Supplementary Material for elaborate discussion).

\section{Conclusion}
In conclusion, we have presented a deterministic protocol to produce a squeezed vacuum state via sequential application of two non-commuting Rabi Hamiltonians of the form $\hat{P}\hat{\sigma}_x$ and $\hat{X}\hat{\sigma}_y$. This interaction can currently be implemented via various methods in trapped-ion and circuit QED systems, but the protocol is not fundamentally limited to those systems and could be relevant for other qubit-oscillator platforms. Unlike previous methods, the protocol deterministically builds the squeezed state through a discrete superposition of coherent states. The protocol does not inherently require qubit measurements, but the performance can be slightly improved by post-selecting on the state of the qubit. The possible amount of quadrature squeezing is ultimately limited by decoherence mechanisms of either the bosonic or qubit modes, while the achievable Fisher information is mainly limited by bosonic noise, and both can be improved by increasing the interaction strength and thus the speed of the protocol.

\begin{acknowledgements}
\section*{Acknowledgements}
This project was supported by the Danish National Re-search Foundation through the Center of Excellence for Macroscopic Quantum States (bigQ). R.F. acknowledges project LTAUSA19099 and 8C20002 from the Ministry
of Education, Youth and Sports of Czech Republic. K.P. acknowledges project 19-19722J of the Grant Agency of Czech Republic (GACR).
\end{acknowledgements}

\section*{References}
\bibliography{References}

%%%%%%%%%%%%%%%%%%%%%%%%%%%%%%%%%%%%%%%%%%%%%%%%%%%%%
%%%%%%%%%%%%%%%%%%%%%%%%%%%%%%%%%%%%%%%%%%%%%%%%%%%%%
% Supplementary
%%%%%%%%%%%%%%%%%%%%%%%%%%%%%%%%%%%%%%%%%%%%%%%%%%%%%
%%%%%%%%%%%%%%%%%%%%%%%%%%%%%%%%%%%%%%%%%%%%%%%%%%%%%

\clearpage
 \setcounter{section}{0}
    \renewcommand{\thesection}{S\arabic{section}}
    \setcounter{equation}{0}
    \renewcommand{\theequation}{S\arabic{equation}}
    \setcounter{figure}{0}
    \renewcommand{\thefigure}{S\arabic{figure}}
\section*{Supplementary Material}

\begin{figure}
\centering
\includegraphics{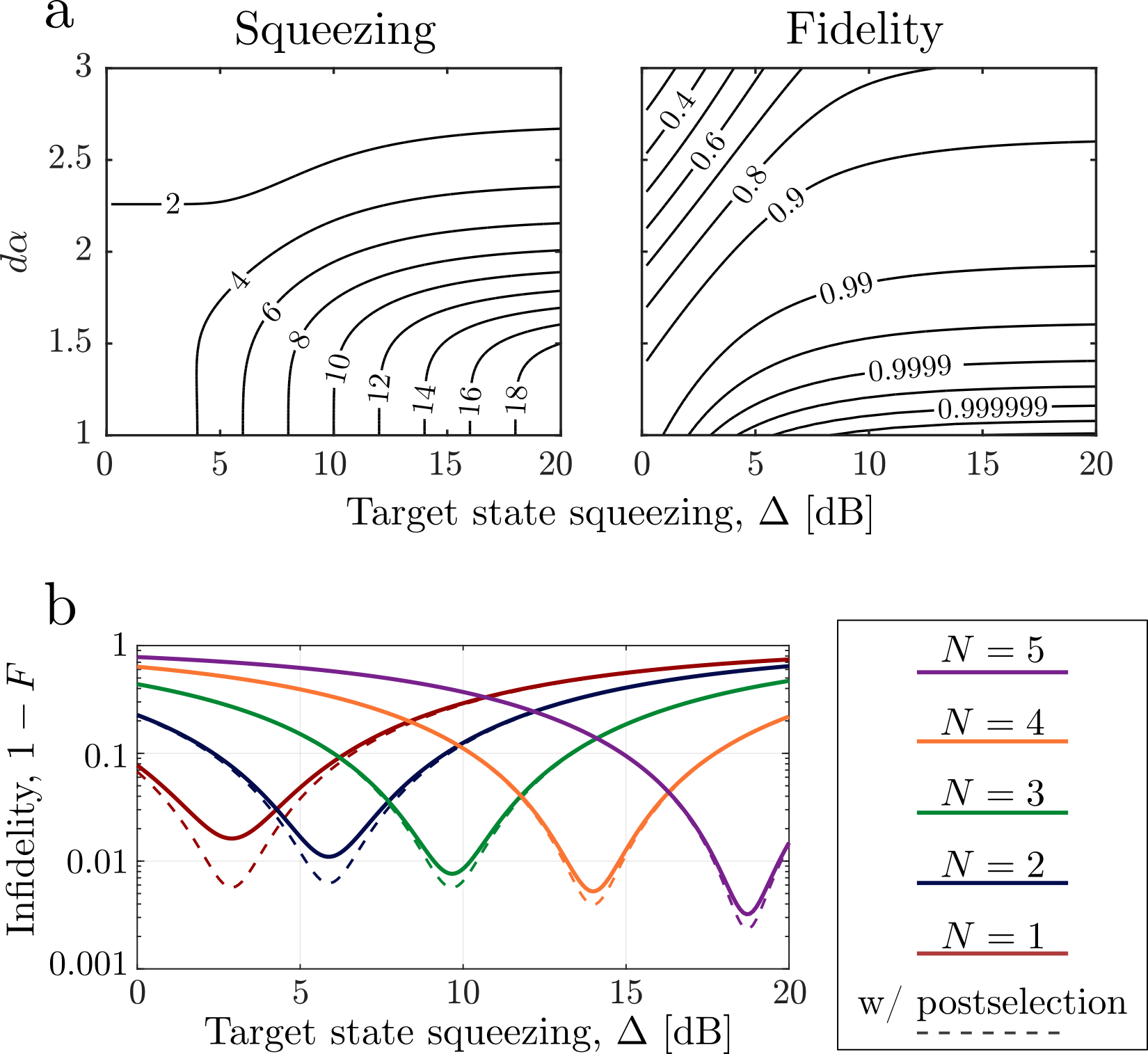}
\caption{(a) Squeezing in dB and fidelity of the coherent state superposition of Eq. \eqref{eq:coherentsuper}. For $d\alpha\lesssim1.5$ the squeezing level matches that of the target squeezed vacuum state, and the fidelity is very high, confirming the approximation of Eq. \eqref{eq:approx}. (b) Infidelity between the state generated by our protocol and a pure squeezed vacuum state with squeezing parameter $\Delta$.}
\label{fig:discrete}
\end{figure}
\begin{figure}
\centering
\includegraphics{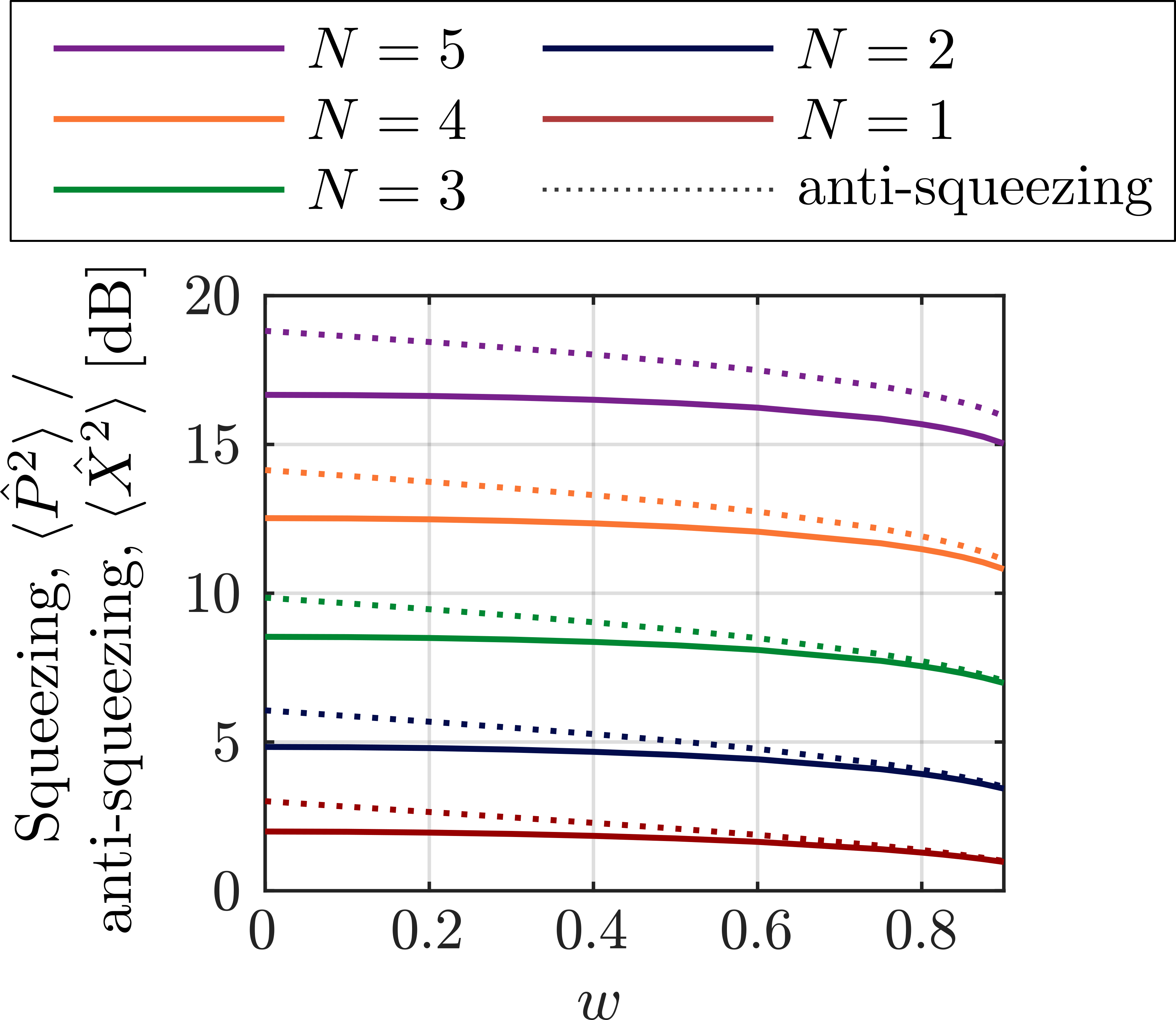}
\caption{Obtainable amount of squeezing and anti-squeezing as a function of the weight parameter $w$ described in the supplementary text \ref{app:anti-squeezing}. By appropriately choosing $w$ one can reduce the excess anti-squeezing without significantly compromising the squeezing.}
\label{fig:anti-squeezing}
\end{figure}
\section{Discrete coherent state representation of squeezed vacuum}\label{app:squeezedApprox}
Here we numerically examine the approximation of Eq. \eqref{eq:approx} of the main text. Specifically, we consider a superposition of equally spaced coherent states with spacing $d\alpha$:
\begin{equation}
    \ket{\psi} = \sum_{s=-\infty}^{\infty}\exp\left(-\frac{(d\alpha (s+1/2))^2}{\Delta^{-2}-1}\right)\ket{d\alpha (s+1/2)} 
    \label{eq:coherentsuper}
\end{equation}
Fig. \ref{fig:discrete}a shows the squeezing of the state, and the fidelity, $|\langle\psi|\textrm{sqvac}\rangle_\Delta|^2$, to the target squeezed vacuum state as a function of $d\alpha$ and $\Delta$. For sufficiently small spacing, $d\alpha\lesssim1.5$, we observe an excellent agreement between the expected and target squeezing level as well as a high fidelity to the target state. Note that for high squeezing, the squeezing levels can be significantly smaller than the target, even if the fidelity is very high, e.g. for $\Delta=20$ dB at $d\alpha=2$. This is because the squeezing level is very sensitive the the small non-Gaussian features which arise from the discretization of Eq. \eqref{eq:coherentsuper}, which is not captured by the fidelity. For this reason we have chosen the squeezing level as the relevant figure of merit throughout this paper. 

In Fig. \ref{fig:discrete}b we show the fidelity of the states generated in our protocol with respect to pure Gaussian squeezed vacuum states. For each $N$ the interaction parameters were chosen to optimize the quadrature squeezing and not the fidelity. Yet, the produced states have fidelities of $F>0.99$, confirming that the generated states are indeed very close to Gaussian squeezed states.

\section{Optimizing excess anti-squeezing}\label{app:anti-squeezing}
In the main text we chose the interaction parameters to optimize the quadrature squeezing, i.e. by minimizing $\langle \hat{P}^2\rangle$. This results in an anti-squeezing a few dB above the transform limit, such that $\langle \hat{P}^2\rangle\langle \hat{X}^2\rangle>\frac{1}{4}$. However, it is possible to reduce the amount of excess anti-squeezing by choosing the interaction parameters slightly differently. For example, one can optimize the function $\langle \hat{P}^2\rangle^{1-w} \left(\langle \hat{P}^2\rangle\langle \hat{X}^2\rangle\right)^w$, where $w$ is a weight parameter determining the relative weight between squeezing and excess anti-squeezing. For $w=0$ we recover the result from the main text. Fig. \ref{fig:anti-squeezing} shows the obtainable amount of squeezing and anti-squeezing as a function of $w$. In conclusion, one can reduce the anti-squeezing to a level close to the transform limit, while only slightly decreasing the squeezing.

\section{Fisher information}
\begin{figure}
\centering
\includegraphics{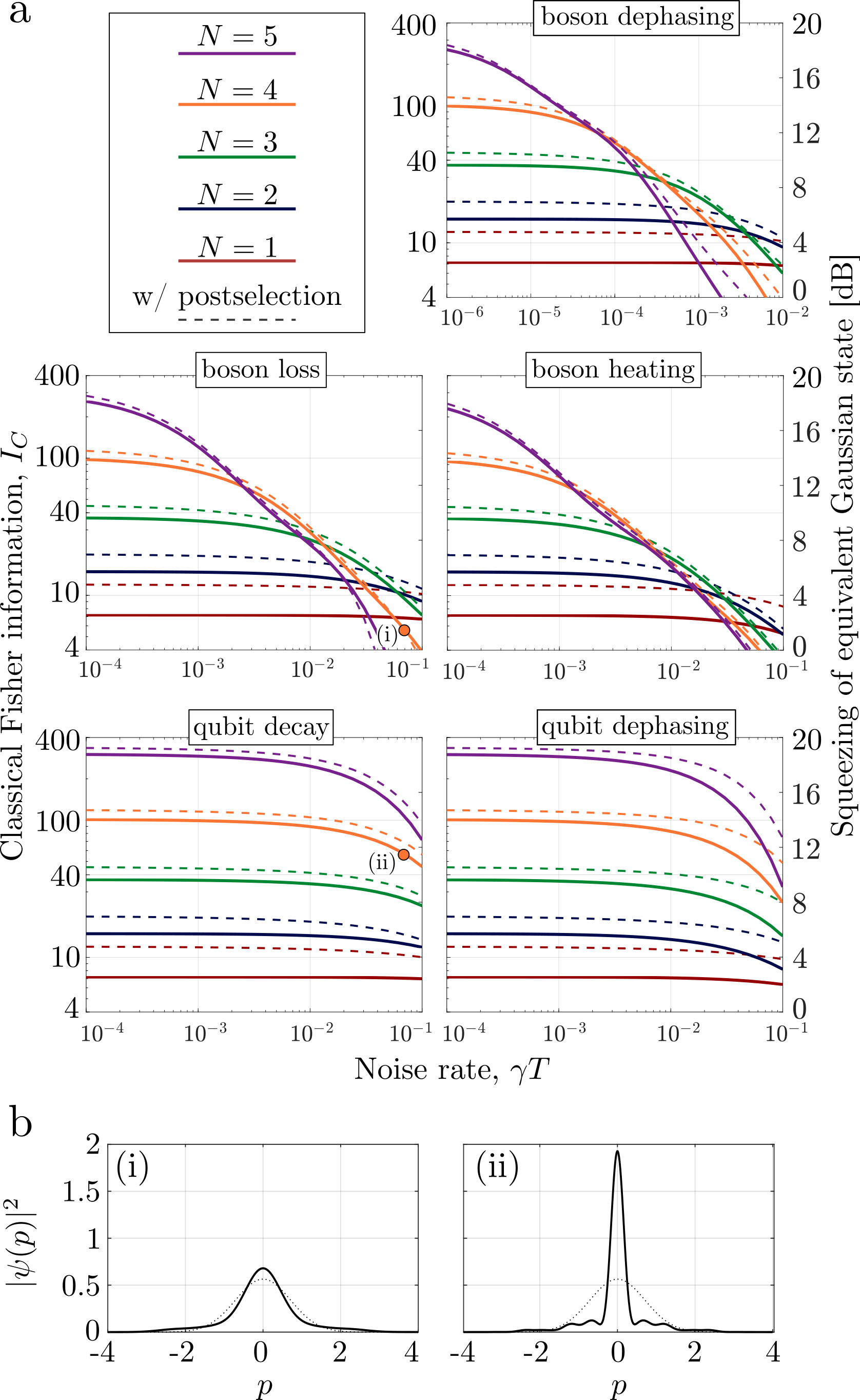}
\caption{(a) Classical Fisher information, $I_C$, as a function of noise rate for various noise sources applied during the protocol. The Fisher information can also be expressed in terms of the amount of squeezing required in a Gaussian state to achieve the same $I_C$, through Eq. \eqref{eq:CFIequiv}, which is shown on the right axis. (b) Example momentum quadrature distributions of the generated states suffering from (i) boson loss and (ii) qubit decay during the preparation protocol, with $\gamma T=7\times 10^{-2}$, as marked in (a). The dotted lines show the quadrature distribution of vacuum for comparison.} 
\label{fig:Fisher}
\end{figure}

Squeezed states are useful for detecting small quadrature displacements with a precision beyond the standard quantum limit set by the vacuum state. In a practical setting, the capability of the state to measure a small momentum displacement, $d$, caused by the displacement operator $\hat{D}(id)=e^{i\sqrt{2}d\hat{X}}$ is given by the classical Fisher information, $I_C$, with respect to homodyne detection \cite{paris2004quantum}:
\begin{equation}
I_C = 2 \int dp \left(\frac{\partial}{\partial p}\log|\psi(p)|^2\right)^2 |\psi(p)|^2,
\end{equation}
where $|\psi(p)|^2=\textrm{Tr}\left(\rho |p\rangle\langle p|\right)$ is the $p$-quadrature probability density. For a Gaussian state the Fisher information is directly related to the quadrature variance as
\begin{equation}
I_C=2/(\langle\hat{P}^2\rangle-\langle\hat{P}\rangle^2), \qquad \textrm{(Gaussian states)} \label{eq:CFIequiv}
\end{equation}
On the other hand, for non-Gaussian states the quadrature variance does not necessarily capture the sensing properties of non-Gaussian states. The underlying steps of our protocol are non-Gaussian, and thus the noise accumulated during the protocol can enhance the non-Gaussian properties of the final state. Fig. \ref{fig:Fisher}a shows the calculated Fisher information of the state prepared by a noisy protocol, similar to Fig. \ref{fig:noise} of the main text. We observe that qubit errors have a significantly smaller impact on the Fisher information compared to the squeezing of Fig. \ref{fig:noise}. Bosonic noise, on the other hand, also impacts the Fisher information, although the Fisher information of the obtained state is generally higher than what would be expected from a Gaussian state with squeezing given by Fig. \ref{fig:noise}. The difference between the bosonic and qubit noise can be understood by looking and the resulting quadrature distribution. Fig. \ref{fig:Fisher}b shows the resulting distributions from a protocol affected by bosonic loss (i) and qubit decay (ii) respectively. While both states have a low amount of squeezing in terms of the variance, the state affected by qubit decay has a much narrower peak, resulting in a higher Fisher information. Thus for sensing applications, the quality of the prepared states is primarily limited by bosonic noise and less by qubit noise.

\end{document}